\documentclass[12pt]{iopart}
\usepackage{epsfig}
\input amssym.def
\input amssym.tex

\renewcommand\({\left(}
\renewcommand\){\right)}

\newcommand\eq[1]{Eq.~(\ref{#1})}

\newcommand\eqreff[1]{(\ref{#1})}

\newcommand\ee{\end{equation}}
\newcommand\be{\begin{equation}}
\newcommand\eea{\end{eqnarray}}
\newcommand\bea{\begin{eqnarray}}

\newcommand\mpl{M_{\rm P}}

\def\calp{{\cal P}}

\newcommand\GeV{\,\mbox{GeV}}

\newcommand\half{^{1/2}}

\begin{document}

\title{$A$-term inflation and the MSSM}

\author{Juan C. Bueno Sanchez, Konstantinos Dimopoulos and David H.~Lyth}
\address{Physics Department, University of Lancaster,
Lancaster LA1 4YB, United Kingdom}
\begin{abstract}
The parameter space for $A$-term inflation is explored with
$W=\lambda_p \phi^p/p\mpl^{p-3}$. With  $p=6$ and $\lambda_p\sim
1$, the  observed spectrum and spectral tilt can be obtained with
soft mass of order $10^2\GeV$ but not with a much higher mass. The
case $p=3$ requires $\lambda_p\sim 10^{-9}$ to $10^{-12}$. The
ratio $m/A$ requires fine-tuning,  which may be justified on
environmental grounds. An extension of the MSSM to include
non-renormalizable terms and/or Dirac neutrino masses might
support either $A$-term inflation or modular inflation.
\end{abstract}

\maketitle

Recently, it has been proposed that the field content of the minimal
supersymmetric standard model (MSSM) describes both the low energy
physics  that is observed at colliders, and inflation after the
observable Universe leaves  the horizon \cite{aegm}. The inflaton is
supposed to correspond to a flat direction. To achieve this, a new
kind of inflation model is formulated, in which the potential is of
the form \be\label{vpot} V
=\frac{1}{2}m^2\phi^2-A\frac{\lambda_p\phi^p}{p\,M_P^{p-3}}
+\lambda_p^2\frac{\phi^{2(p-1)}}{M_P^{2(p-3)}} \ee The last term
corresponds to a minimal globally supersymmetric theory with
superpotential \be W=\frac{\lambda_p}p \phi^p /\mpl^{p-3}. \ee The
first two terms are soft supersymmetry breaking parameters.
Supersymmetry-breaking is taken to be gravity-mediated so that
$A\sim m$. For parameter values that allow inflation, $m\gg H$ where
$H$ is the Hubble parameter, and as a result the phase of $\phi$
adjusts to minimize the potential, giving \eq{vpot}.

This  kind of inflation model is  further explored in \cite{p06},
where it is called  $A$-term inflation. The model  is interesting
in its own right, and might apply with $\phi$  a gauge singlet, or
a flat direction of any gauge group. In this note, the parameter
space for $A$-term inflation is explored, and a  fine-tuning issue
is addressed.

The model works because the parameters can be chosen so that
$V^{\prime}$ and $V^{\prime\prime}$ are very small at some point,
allowing inflation to take place near that point. Indeed,
$V^\prime=V^{\prime\prime}=0$ exactly at the point
\begin{equation}\label{phi0}
  \phi_0=\left(\frac{mM_P^{p-3}}{\lambda_p\sqrt{2p-2}}\right)^{1/(p-2)},
\end{equation}
provided that \be \label{m2} m^2= \frac{A^2}{8(p-1)} . \ee More
generally, inflation can occur near $\phi_0$ if \be
\label{m2delta} m^2= \frac{A^2}{8(p-1)} \( 1 + \frac{\delta^2}
{m^2} \) \label{con2} \ee with the parameter $\delta^2/m^2$
sufficiently small. To first order in this parameter, $\delta^2$
is the shift in $m^2$ away from the value \eqreff{m2}, at fixed
$A$. If the parameter is positive then $V^\prime$ is positive for
all $\phi$. If it is  negative, $V$ has a maximum to the left of
$\phi_0$, and viable inflation can take place only to the left of
the maximum.

This model is stable  against loop corrections, supergravity
corrections and  corrections from higher-order terms in the
tree-level potential. Indeed \cite{p06}, such corrections just
multiply the right hand side of \eq{m2} (hence of \eq{m2delta}) by
some  factor $(1+f)$ with $|f|\ll 1$, and the right hand side of
\eq{phi0} by some  factor $(1+g)$ with $|g|\ll 1$. These factors
have a negligible effect on the predictions and would be invisible
on our plots, and so we ignore them in the following.

For a given $p$, the parameters of the model are $\lambda_p$,
$m^2$ and $\delta^2$. Assuming that the inflaton perturbation
generates the curvature perturbation, constraints on the model are
provided by the observed values of the spectrum of the curvature
perturbation and of spectral index: \bea
\calp_{\zeta}\half &=&  4.8\times 10^{-5} \\
n &=& 0.948 \pm 0.015 \eea The uncertainty in the spectrum is
negligible for the present purpose, and the uncertainty on $n$ is
the current 1-$\sigma$ value \cite{wmap3} for models (like the
present one) which give a negligible tensor perturbation.

These constraints are evaluated analytically in the Appendix. They
depend on the number $N$  of $e$-folds of inflation after the
observable  Universe leaves the horizon, and we have set $N=50$
which corresponds to continuous radiation domination between  the
end of inflation and the present matter-dominated era. That
situation is expected  because the flat direction is expected to
quickly decay and thermalize at the end of inflation \cite{os}.
Delayed reheating would not alter $N$ much, but one or two bouts
of thermal inflation \cite{thermal} could reduce it  by 10 or 20,
which as we see would significantly alter the constraints.

\begin{figure}[htbp]\hspace{1cm}
\epsfig{file=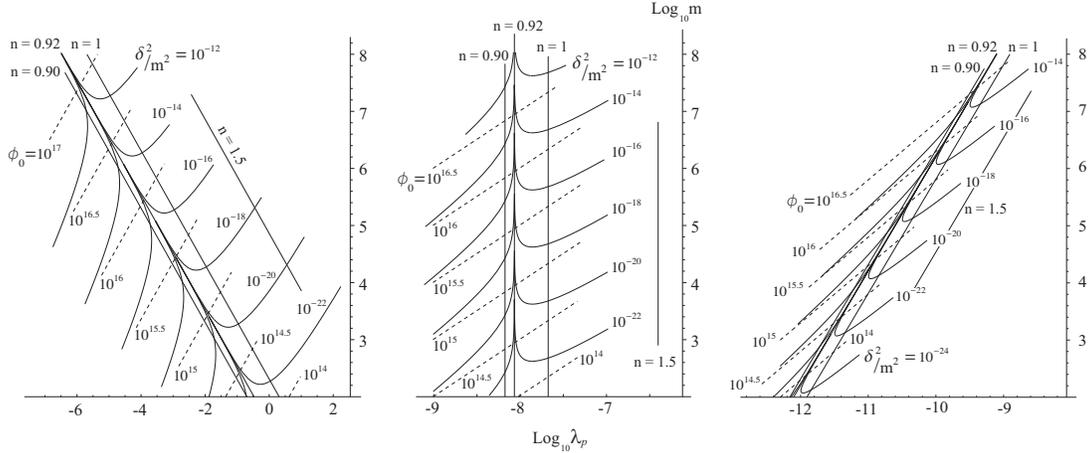,width=14.5cm}\caption{\label{mvsl}The
graphic shows lines of constant $n$ (solid line), $\phi_0$ (dashed
line) and $\delta^2/m^2$ (thick solid line) for the cases $p=6$
(left hand panel), $p=4$ (central panel) and $p=3$ (right hand
panel), as obtained from Eqs.~(\ref{phi0}), (\ref{delta^2}) and
(\ref{phi0^2}), in the $m$-$\lambda_p$ plane ($\phi_0$ and $m$ are
expressed in \GeV). Also shown the region where the spectral index
is within the observational limits, along with lines $n=1.5$ and
$n=0.92$ (corresponding to $\delta^2/m^2=0$) added for reference.
The amount of fine-tuning $\delta^2/m^2$ ranges between
$\sim10^{-22}$ and $\sim10^{-12}$.}
\end{figure}

The constraints are plotted in Figure \ref{mvsl}, in which lines
of constant $n$, $\phi_0$ and $\delta^2/m^2$ are shown. For each
point in the $m$-$\lambda_p$ plane, we choose $\delta^2$ to give
the observed spectrum. The  panels, from left to right, correspond
to $p=6$, $p=4$ and $p=3$, respectively.

We discuss first the $p=6$ results. If the ultra-violet cutoff of
the theory is $\mpl$ one generally expects $\lambda_p$ roughly of
order 1, or perhaps of order $1/p!\sim 10^{-3}$. If instead there
is a GUT corresponding to a cutoff $10^{-2}\mpl$ these expected
values are increased by a factor $10^{2(p-3)}=10^{6}$. The MSSM
places $m$ in the range $10^2$ to $10^3\GeV$.

Our plot confirms the remarkable finding of \cite{aegm}, that the
spectrum and the spectral index can be consistent with observation
for values of $\lambda_p$ and $m$ within the expected range. In
particular, $\delta^2=0$ gives $n=1-4/N$. With $N=50$ this is a
bit low compared with observation, though  reducing $N$ by 20 or
so would bring it up to the central value.

Our plot shows two things which were not anticipated. First, the
ratio $m^2/A^2$ must be fine-tuned to the value \eqreff{m2} with
extraordinary accuracy of order $10^{-20}$, in order to reproduce
the observed spectrum.\footnote {It was pointed out in \cite{p06}
that inflation {\it per se} requires the value to be tuned with
accuracy of order $10^{-4}$, but observational constraints were not
considered there.} Second, a further modest fine-tuning is needed to
get $n$ within the range $0.90$ to $1.00$, which observation surely
requires. These fine-tunings make $A$-term inflation very different
from inflation with potentials of the usual form
\cite{treview,book,laila}. For those potentials the predictions are
not very sensitive to the parameters (though one should remember
that the potentials are usually obtained by setting equal to zero
parameters that might reasonably have been expected to be
significant). Also,  $n$ is automatically  a bit below 1 if $\ln V$
is strongly concave-downwards as is the case for several
well-motivated shapes of the  potential.

Our plot shows that the constraints cannot be satisfied (with
$\lambda_p$ in the expected range) if $m$ is increased by several
orders of magnitude. 
However, 
this is consistent with the observed normalization of the spectrum;
it is the high value of $n$ that is inconsistent.

What about the fine-tuning? In recent years, there has been
increasing interest in the  idea that there exists a landscape of
values, for  parameters which would formerly have been regarded as
fixed. The landscape is supposed to  correspond  to possible
solutions of the equations of a fundamental theory, which might
perhaps be realised somewhere in  the universe. The fundamental
theory is usually supposed to be string theory \cite{strings},  or
in some cases just field theory.

The landscape might allow fine-tuning to be justified on
environmental (anthropic) grounds. In the context of cosmology,
the landscape was proposed to allow fine-tuning of the axion
misalignment angle \cite{linde}, of the cosmological constant
\cite{weinberg} or of the value of a curvaton field
\cite{curvaton}. In the context of particle physics, it might
explain the one-percent fine-tuning of the MSSM \cite{mssm}, or
the drastic fine-tuning of Split Supersymmetry \cite{split} or no
supersymmetry at all \cite{nosusy}. Let us see how the landscape
might justify the fine-tuning of $A$-term inflation.

\begin{figure}[htbp]\hspace{1.5cm}
\epsfig{file=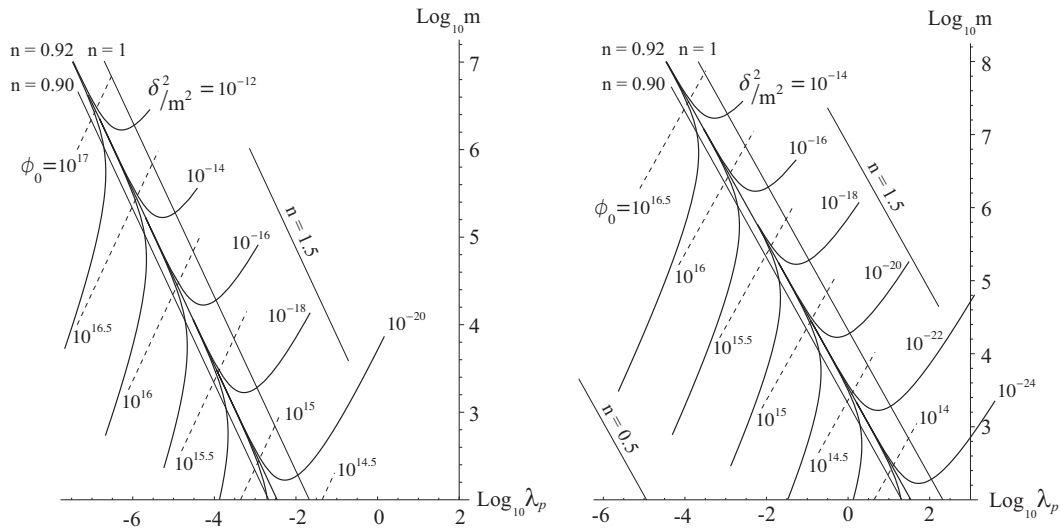,width=14cm}\caption{\label{antr}The graphic
shows lines of constant $n$, $\phi_0$ and $\delta^2/m^2$ in the
$m$-$\lambda_p$ plane. The left hand panel corresponds to
$\calp_{\zeta}^{1/2}$ ten times smaller than the observed value and
the right hand panel to $\calp_{\zeta}^{1/2}$ ten times bigger. When
$\calp_{\zeta}^{1/2}$ is reduced below its observed value,
$V^{\prime}$ must increase in order to reproduce the observed
spectrum. This, in turn, results in an increase in $\delta^2/m^2$.
On the contrary, when $\calp_{\zeta}^{1/2}$ increases $\delta^2/m^2$
decreases. Given that $(\delta^2/m^2)/(\phi_0/M_P)^4$ does not
depend on the particular value $\calp_{\zeta}^{1/2}$ the relative
position of the lines of constant $\delta^2$ and $\phi_0$ remains
unchanged, as can be seen by comparing the left hand and right hand
panels.}
\end{figure}

The possibility arises because environmental considerations
 require that $\calp_{\zeta}\half$  be  within a factor 10 or
so of its observed value \cite{tr}. The predictions for
$\calp_{\zeta}\half$ at these limits are shown in Figure
\ref{antr}. To get a feel for what is going on, consider the
values $m=10^3\GeV$ and $\delta^2=0$ with $\lambda_p$ chosen to
give the observed value of $\calp_{\zeta}\half$. Now allow
$\delta^2$ to vary keeping  $m$ and $\lambda_p$ fixed; we see that
requiring
 $\calp_{\zeta}\half$  to be within its anthropic range requires
$|\delta^2/m^2|\lesssim 10^{-20}$. The situation is similar for
other choices of $m$ and $\lambda_p$.

Of course, to  allow a landscape of values for $m$ raises the
question of why it has the low value corresponding to the MSSM. %
(As we already remarked, the low value is essential so that the
spectral index has a chance of being compatible with observation.)
Interestingly, the question may possibly be answered by considering
the initial condition for inflation. For negative $\delta^2$,
corresponding to $n<0.92$ and low $m^2$, inflation with this
potential can start with eternal inflation near the maximum
\cite{eternal}. The indefinite duration of eternal inflation
arguably justifies such an initial condition. For positive
$\delta^2$, we see in the Appendix that there is
 no regime of eternal
inflation  except for a negligible interval
(which does not allow the spectral index to become significantly
different above  $n=0.92$). The initial condition may still be
justified in that case if, as is quite reasonable, there is an
early phase of primordial inflation at the usual high scale
through one of the usual mechanisms, but with the parameters such
that the curvature perturbation generated then is much less than
the observed one.

Indeed, after such inflation $\phi$ may take on a range of values
within the inflated patch, but the environmental constrain on
$\calp_\zeta$ means that we can live only in selected regions. (Such
inflation may also begin with eternal inflation, which arguably
makes it irrelevant whether $A$-term inflation supports eternal
inflation.) Now comes the crucial point; as $m^2$ increases at fixed
$\lambda_p$, the viable range of initial values of $\phi$ becomes
smaller, and so does the possible amount of inflation. Arguably,
this means that the probability that  we live in a region with such
a value decreases. Hence low values of $m^2$ may be favoured.
Finally, values below $10^2\GeV$ or so may be disfavoured since the
MSSM does not then reproduced the SM as a good  approximation.
Provided that $\lambda_p$ is fairly close to 1, the combined effect
of these arguments is to favour $m\sim 10^2\GeV$.

Instead of fixing $\lambda_p$, we might consider a landscape of
values for both it and $m$. Then the viable range of initial
values of the inflaton field, and the total amount of inflation,
increases as we move down and to the left in the plot of Figure
\ref{antr}. We lose the environmental argument for low $m$ in that
case.

Now we move on to the case $p=3$. There is no factor $\mpl$ in
\eq{vpot} so that $\lambda_p$ is a renormalizable coupling. Figure
\ref{mvsl} confirms the finding of \cite{akm}, that the
observational constraints require $\lambda_p\sim 10^{-12}$ if  $m$
is of order $10^2$ to $10^3\GeV$. As was pointed out there, the
small $\lambda_p$ is the one required to generate a Dirac neutrino
mass of the observed value, if $\phi$ is a suitable flat direction
of the MSSM extended to make the neutrino field a Dirac field. The
fine-tuning is about the same as in the $p=6$ case. Now the viable
range of initial values of the inflaton field, and the total amount
of inflation, increases as we move up and to the left in the
corresponding plot of Figure \ref{mvsl}. If we could find a reason
for favouring $m\sim 10^2\GeV$, the value  $\lambda_p$ would be
anthropically favoured, and so would the observed value of the
neutrino mass in the model of \cite{akm}.

This discussion of small $\lambda_p$ in the case $p=3$ reminds us,
somewhat uncomfortably, that the received wisdom of expecting
$\lambda_p\sim 1$ for $p>3$  lacks  a  firm  foundation. In the
case of $p=3$ couplings, one is forced to accept a wide range of
values to explain the wide range of particle masses generated by
the Higgs mechanism (down to $10^{-5}$ or so for the coupling that
determines the electron mass). There is no generally accepted
scheme for explaining this range, and none for justifying the
 expectation $\lambda_p\sim 1$ for $p>3$ which is really made just on
grounds of simplicity.

In fact, a completely different view about these coefficients is
also quite natural; that the potential in a flat direction is of the
form $V=V_0f(\phi/\mpl)$ with $f$ and its low derivatives having
magnitude of order 1 at a typical point in the interval
$0<\phi\lesssim \mpl$. This possibility is mentioned in \cite{drt}
in the context of Affleck-Dine baryogenesis. In the context of
string theory it would correspond to $\phi$ being a modulus with the
origin a point of enhanced symmetry \cite{polchinskybook}. A typical
term in the power series expansion of $V$ is now of order $\pm V_0
(\phi/\mpl)^n$, and all terms are important at $\phi\sim\mpl$. The
mass is of order $V_0\half/\mpl$.

Inflation with this type of potential \cite{andrei83} is usually
called modular inflation. The inflaton is usually supposed to be a
gauge singlet, but as was noticed in \cite{p06} it could as well
be a flat direction of a gauge multiplet. The fundamental
assumption for modular inflation is that the potential has a
maximum at $\phi\sim\mpl$. (Given the form of $f$ there could
hardly be a maximum at $\phi\ll\mpl$, but of course that does not
mean that there must be a maximum at $\phi\sim\mpl$.)

At the maximum, the slow roll parameter $\eta$ is equal to $f''$.
This would typically be of order $-1$ which would correspond to
fast-roll inflation \cite{fast}, which  is  viable only if the
curvature perturbation is generated from the vacuum fluctuation of a
field different from the inflaton \cite{dl}. The idea of modular
inflation is that one gets lucky, so that at the maximum
$|f''|\lesssim 10^{-2}$. Then one can expect
\cite{treview,book,laila} the power-series expansion of $V$ around
the maximum to be dominated by higher powers, giving an effective
potential $V\simeq V_0[ 1- (\phi/\mu)^p ]$ with $p\gtrsim 3$ not
necessarily an integer. This allows the curvature perturbation to be
generated from the vacuum fluctuation of the inflaton if $V_0$ is of
order $(10^{15}\GeV)^4$, corresponding to $m\sim 10^{12}\GeV$, and
$n$ is then automatically within the observed range. We conclude
that inflation with an MSSM flat direction corresponding to a
modulus can generate the observed curvature perturbation if the mass
is large corresponding to Split Supersymmetry. (In contrast A-term
inflation is not viable in the context of Split Supersymmetry since
an A term is forbidden in that case by observational constraints.)

Returning to $A$-term inflation, we have dealt so far  with the
cases $p=6$ and $p=3$ that alone are possible within the MSSM. As
seen in Figure \ref{antr}, the case $p=4$ satisfies the
observational constraints only for a unique coupling
$\lambda_p\simeq 10^{-8}$. The intermediate case $p=5$ (not shown)
is similar to $p=6$ but the slope of the allowed regime is steeper
and corresponds to smaller $\lambda_p$. We have not considered cases
$p>6$, because as $p$ increases the situation that the term with
that power actually dominates begins to look increasingly unlikely,
owing to the extraordinarily strong suppression of lower powers that
this  requires \cite{gkm}.

In conclusion, we have explored the parameter space for $A$-term
inflation, and have identified a severe fine-tuning. We have
presented some environmental considerations that may help to
justify this tuning. While  recognising that such  arguments are
extremely controversial,
 we feel that they are worthy of  consideration now that the idea of a
landscape is under active discussion.

By way of closing, we would like to remind the reader of a
fundamental point. For  $A$-term inflation  with $p>3$, as well as
for modular inflation, the potential involves non-renormalizable
terms. Such terms are negligible in the low-energy theory
describing astrophysical and terrestrial processes. This means
that  $A$-term or modular inflation  using a flat direction of the
MSSM cannot be tested in the laboratory; such models are invoking
an extension of the MSSM, not to include more fields but to
include more interactions. Only the proposal of \cite{akm} can be
so tested, through its assertion that the neutrino is a Dirac
particle.

\ack The authors wish to thank Anupam Mazumdar and Rouzbeh
Allahverdi for useful discussions. DHL is supported by PPARC grant
  PPA/Y/S/2002/00272   and EU grant MRTN-CT-2004-503369.

\appendix
\setcounter{section}{1}
\section*{Appendix}
\subsection{Constraints from CMB normalization}

Near $\phi=\phi_0$ we may write
\begin{equation}
  V(\phi)=V(\phi_0)+\frac{1}{6}V^{\prime\prime\prime}(\phi_0)(\phi-\phi_0)^3\,,
\end{equation}
where $V^{\prime\prime\prime}(\phi_0)=2(p-2)^2\frac{m^2}{\phi_0}$
and $V(\phi_0)=\frac{(p-2)^2}{2p(p-1)}m^2\phi_0^2$. If $m^2$ is
reduced by a small amount $\delta^2\equiv\delta(m^2)$ the
potential becomes
\begin{equation}
  V\simeq
  V(\phi_0)+\frac{1}{6}V^{\prime\prime\prime}(\phi_0)\chi^3+\delta^2\phi_0\chi\,,
\end{equation}
where we have introduced the field $\chi=\phi-\phi_0$. In order
for the flat direction to be responsible for the observed
curvature perturbation we impose the CMB normalization
\begin{equation}
  \calp_{\zeta}^{1/2}=\frac{1}{\sqrt{12\pi^2}}\frac{V^{3/2}}{M_P^3V^{\prime}(\chi_{\ast})}\,,
\end{equation}
which at $p$ and $m$ fixed determines the field value $\chi_*$
when the observable Universe leaves the horizon in terms of
$\delta^2$ and $\phi_0$. We obtain
  \begin{equation}\label{Xast}
  \frac{|\chi_{\ast}|}{\phi_0}=\left[\frac{1}{\calp_{\zeta}^{1/2}\sqrt{12\pi^2}}\frac{p-2}{(2p(p-1))^{3/2}}
  \frac{m}{M_P}\left(\frac{\phi_0}{M_P}\right)^2-\frac{1}{(p-2)^2}\frac{\delta^2}{m^2}
  \right]^{1/2}\,.
\end{equation}
In the slow-roll approximation, the number of $e$-foldings $N$
after the observable Universe leaves the horizon is given
by\footnote{Here we neglect the contribution from $\chi$ at the
end of inflation, $\chi_e$. This approximation holds until
$|\eta|\sim1$ and amounts to taking the limit
$|\chi_e|\to\infty$.}
\begin{equation}\label{efolds}
  \frac{1}{p-2}\sqrt{\frac{\delta^2}{m^2}}\cot\left[\frac{2p(p-1)}{p-2}N
  \left(\frac{M_P}{\phi_0}\right)^2\sqrt{\frac{\delta^2}{m^2}}\right]
=\frac{-\chi_*}{\phi_0}\,.
\end{equation}
When $\delta^2$ becomes negative the corresponding equation can be
readily found by analytic continuation.

Once a particular value $\lambda_p$ is selected, and having fixed
$m$ and $p$, there is only one possible value $\delta^2$ for which
the flat direction generates the observed spectrum of
perturbations.

In order to solve \eq{efolds} for $\delta^2$ and $\phi_0$ we
introduce the variables
\begin{equation}
  \Delta^2\equiv\frac{\delta^2}{m^2}\left(\frac{M_P}{\phi_0}\right)^4\quad{\rm
  and}\quad\varphi\equiv \left(\frac{\phi_0}{M_P}\right)^2\,,
\end{equation}
where it should be noted that $\Delta^2$ carries the same sign as
$\delta^2$. In these variables \eq{efolds} leads to a
parametrisation of $\varphi$ in terms of $\Delta^2$. Using then
\eq{efolds}, written as $\varphi=\varphi(\Delta^2)$, the spectral
index $n\approx1+2\eta$ becomes
\begin{equation}\label{spind}
  n=1-\frac{8p(p-1)}{p-2}
  \sqrt{\Delta^2}
  \,\cot\left[\frac{2p(p-1)}{p-2}N\sqrt{\Delta^2}\right]\,,
\end{equation}
\begin{figure}[htbp]
\centering
\epsfig{file=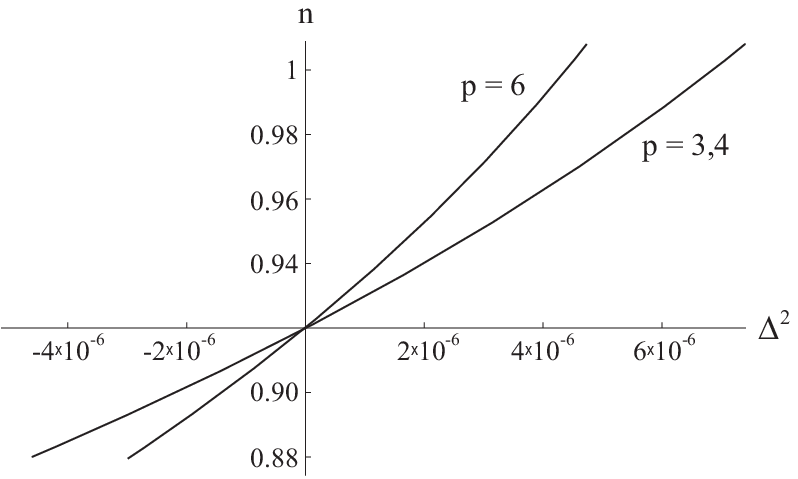,width=8.02cm}\caption{\label{fig:si1}Spectral
index $n$ for $p=3,4$ and $p=6$ taking $N=50$. The curves $p=3$ and
$p=4$ are indistinguishable.}
\end{figure}
which allows us to parametrise implicitly
$\Delta^2=\Delta^2(n,N,p)$. With this we can use
$\varphi=\varphi(\Delta^2)$ to obtain $\delta^2$ and $\phi_0$
\begin{equation}\label{delta^2}
  \frac{\delta^2}{m^2}=\Delta^2\varphi^2=\left(\frac{(p-2)^3}{(2p(p-1))^{3/2}}\frac{1}
  {\calp_{\zeta}^{1/2}\sqrt{12\pi^2}}\frac{m}{M_P}\right)^2\frac{1}{\Delta^2}
  \sin^4\left[\frac{2p(p-1)}{p-2}N\sqrt{\Delta^2}\right]
\end{equation}

\begin{equation}\label{phi0^2}
\left(\frac{\phi_0}{M_P}\right)^2=\varphi=\left(\frac{(p-2)^3}{(2p(p-1))^{3/2}}\frac{1}
{\calp_{\zeta}^{1/2}\sqrt{12\pi^2}}\frac{m}{M_P}\right)\frac{1}{\Delta^2}
\sin^2\left[\frac{2p(p-1)}{p-2}N\sqrt{\Delta^2}\right]\,.
\end{equation}
When $\Delta^2<0$ it follows that $\delta^2$ has the same sign as
$\Delta^2$ and that $\phi_0$ is always positive, as they must.

\subsection{Eternal inflation?}
Now we determine the condition for the Universe to undergo a phase
of eternal inflation when $\delta^2$ is positive, thus not
resulting in the formation of any extrema in the potential. A
region of the Universe undergoes eternal inflation if
\begin{equation}\label{etinf}
  |V^{\prime}|< \frac{3}{2\pi}H^3
\end{equation}
in it. When the $A$-term does not generate a local minimum
$V^{\prime}$ finds its minimum value at $\phi_0$:
$V^{\prime}(\phi_0)=\phi_0\delta^2$. In this case \eq{etinf} turns
into
\begin{equation}
  \frac{\delta^2}{m^2}<\frac{3}{2\pi}\left(\frac{(p-2)^2}{6p(p-1)}\right)^{3/2}
  \!\!\frac{m}{M_P}\left(\frac{\phi_0}{M_P}\right)^2\,.
\end{equation}
Using now Eqs.~(\ref{spind}), (\ref{delta^2}) and (\ref{phi0^2}) the
equation above becomes
\begin{equation}
  \sin^2\left[\frac{2p(p-1)}{p-2}N
\sqrt{\Delta^2}\right]<\calp_{\zeta}^{1/2}\,.
\end{equation}
Keeping the first term in a series expansion, this implies
$\Delta^2<{\cal O}(10^{-10})$ for the cases $p=3,4$ and $p=6$,
which in view of Fig. \ref{fig:si1} can only be satisfied for
$n\approx0.92$.

\end{document}